\documentclass{siamltex}

\RequirePackage{amsmath}

\newif\ifsiam
\siamtrue

\ifsiam\else
\RequirePackage{amsthm}
\fi

\long\def\comment#1{}

\ifsiam
\newtheorem{fact}[theorem]{Fact}

\def\squarebox#1{\hbox to #1{\hfill\vbox to #1{\vfill}}}
\newcommand{\qed}{\vbox{\hrule\hbox{\vrule\squarebox{.667em}\vrule}\hrule}}

\newenvironment{remark}{\noindent 
{\sc Remark:}}{\nolinebreak\hspace{2ex}\nolinebreak\qed \par}\else

\theoremstyle{plain}
\newtheorem{theorem}{Theorem}[section]
\newtheorem{lemma}[theorem]{Lemma}

\newtheorem{corollary}[theorem]{Corollary}

\newtheorem{fact}[theorem]{Fact}

\theoremstyle{definition}

\newtheorem{definition}[theorem]{Definition}

\theoremstyle{remark}
\newtheorem*{remark}{Remark}

\fi

\makeatletter

\def\labelprop#1#2{%
    \@bsphack
    \if@filesw
        {\let\thepage\relax
         \def\protect{\noexpand\noexpand\noexpand}%
         \edef\@tempa{%
            \write\@auxout{\string\newlabel{#2}{{#1~\@currentlabel}{\thepage}}}%
         }%
         \expandafter}\@tempa
        \if@nobreak \ifvmode\nobreak\fi\fi
    \fi
    \@esphack
}

\makeatother

\newcommand{\refprop}[1]{\ref{#1}}

\def\set#1{ \{ #1 \} }

\newcommand{\ceil}[1]{\left \lceil #1 \right\rceil}

\def\NP{\ensuremath{{\cal N \cal P}}}

\def\SP{\ensuremath{\sharp {\cal P}}}

\def\Olog{\tilde{O}}

\newcommand{\dee}[1]{{\mathop{\null\mbox{\boldmath $ \,d$}}} #1}

\def\ie{{\it i.e.}}

\newlength{\figtxtwid}
\long\gdef\boxitnew#1{\dimen200 = \hsize \advance\dimen200 by -7pt
\begingroup\vbox{\hrule \hbox to \hsize{\vrule\kern3pt
      \vbox{\hsize \dimen200 \kern3pt#1\kern3pt}\hfil \kern3pt\vrule}\hrule}\endgroup}
 
\newlength{\myfigwidth}
\newif\ifdraft
\drafttrue
\newif\ifoldversion
\oldversionfalse

\AtBeginDocument{
\ifdraft
\typeout{WARNING: margin notes visible}
\fi
\ifoldversion
\typeout{WARNING: generating old version}
\fi
}

\newenvironment{proofof}[1]{\par\noindent{\bf Proof of #1:}}{\qed} 

\comment{===DEFINED IN PROC.STY
}

\def\fail{\mbox{\rm FAIL}}
\def\rel{\mbox{\rm REL}}

\begin{document}

\hyphenation{FPRAS}

\title{A Randomized Fully Polynomial Time Approximation Scheme for the All
Terminal Network Reliability Problem}

\author{David R. Karger\thanks{\sloppy MIT Laboratory for Computer Science,
Cambridge, MA 02138.   Parts of this work were done at AT\&T Bell
Laboratories. \protect\newline email: 
{\tt karger@lcs.mit.edu.} URL: {\tt http://theory.lcs.mit.edu/\~\/karger}}}

\maketitle

\begin{abstract}
  The classic all-terminal network reliability problem posits a graph,
  each of whose edges fails independently with some given probability.
  The goal is to determine the probability that the network becomes
  disconnected due to edge failures.  This problem has obvious
  applications in the design of communication networks.  Since the
  problem is $\SP$-complete, and thus believed hard to solve exactly,
  a great deal of research has been devoted to {\em estimating} the
  failure probability.  In this paper, we give a {\em fully polynomial
  randomized approximation scheme} that, given any $n$-vertex graph
  with specified failure probabilities, computes in time polynomial in
  $n$ and $1/\epsilon$ an estimate for the failure probability that is
  accurate to within a relative error of $1\pm\epsilon$ with high
  probability.  We also give a deterministic polynomial approximation
  scheme for the case of small failure probabilities.  Some extensions
  to evaluating probabilities of $k$-connectivity, strong connectivity
  in directed Eulerian graphs, $r$-way disconnection, and evaluating
  the Tutte Polynomial are also described.
\end{abstract}

\section{Introduction}


\subsection{The Problem}

We consider a classic problem in reliability theory: given a network
on $n$ vertices, each of whose $m$ links is assumed to fail (disappear)
independently with some probability, determine the probability that
the surviving network is connected.  The practical applications of
this question to communication networks are obvious, and the problem
has therefore been the subject of a great deal of study.  A
comprehensive survey can be found in~\cite{Colbourn}.

Formally, a network is modeled as a graph $G$, each of whose edges $e$
is presumed to fail (disappear) with some probability $p_e$, and thus
to survive with probability $q_e=1-p_e$.  Network reliability problems are
concerned with determining the probabilities of certain
connectivity-related events in this network.  The most basic question
of {\em all-terminal network reliability} is determining the
probability that the network stays connected.  Others include
determining the probability that two particular nodes stay connected
(two terminal reliability), and so on.

Most such problems, including the two just mentioned, are
$\SP$-complete~\cite{Valiant:SharpP,Provan:Reliability}.  That is,
they are universal for a complexity class at least as intractable as $\NP$ and
therefore seem unlikely to have polynomial time solutions.  Attention
therefore turned to approximation algorithms.  Provan and
Ball~\cite{Provan:Reliability} proved that it is $\SP$-complete even
to {\em approximate} the reliability of a network to within a relative
error of $\epsilon$.  However, they posited that the approximation
parameter $\epsilon$ is part of the input, and used an exponentially
small $\epsilon$ (which can be represented in $O(n)$ input bits) to
prove their claim.  They note at the end of their article that ``a
seemingly more difficult unsolved problem involves the case where
$\epsilon$ is constant, \ie\ is not allowed to vary as part of the
input list.''

Their idea is formalized in the definition of a {\em polynomial
  approximation scheme} (PAS).  In this definition, the performance
measure is the running time of the approximation algorithm as a
function of the problem size $n$ and the error parameter $\epsilon$,
and the goal is for a running time that is polynomial in $n$ for each
fixed $\epsilon$ (e.g. $2^{1/\epsilon}n$).  If the running time is
also polynomial in $1/\epsilon$, the algorithm is called a {\em fully
  polynomial approximation scheme (FPAS)}.  An alternative
interpretation of an FPAS is that it has a running time polynomial in
the input size when $\epsilon$ is constrained to be input in unary
rather than binary notation.  When randomization is used in an
approximation scheme, we refer to a {\em polynomial randomized
  approximation scheme (PRAS)} or {\em fully polynomial randomized
  approximation scheme (FPRAS)}.  Such algorithms are required to
provide an $\epsilon$-approximation with probability at least $3/4$;
this probability of success can be increased significantly (e.g., to
$1-1/n$ or even $1-1/2^n$) by repeating the algorithm a small number of
times~\cite{Motwani:RandomizedAlgorithms}.

Deterministic FPASs for nontrivial problems seem to be quite rare.
However, FPRASs have been given for several $\SP$-complete problems
such as counting maximum matchings in dense graphs~\cite{JS:Matching},
measuring the volume of a convex polytope~\cite{Dyer:Volume}, and {\em
  disjunctive normal form (DNF) counting}---estimating the probability
that a given DNF formula evaluates to true if the variables are made
true or false at random~\cite{Karp:Reliability}.  In a plenary talk,
Kannan~\cite{Kannan:Counting} raised the problem of network
reliability as one of the foremost remaining open problems needing an
approximation scheme.

\subsection{Our Results}

In this paper, we provide an FPRAS for the all-terminal network
reliability problem.  Given a failure probability $p$ for the edges,
our algorithm, in time polynomial in $n$ and $1/\epsilon$, returns a
number $P$ that estimates the probability $\fail(p)$ that the graph
becomes disconnected.  With high probability,\footnote{The phrase {\em
    with high probability} means that the probability it does not
  happen can be made $O(n^{-d})$ for any desired constant $d$ by
  suitable choice of other constants (typically hidden in the
  asymptotic notation).} $P$ is in the range $(1\pm\epsilon)\fail(p)$.
The algorithm is Monte Carlo, meaning that the approximation is
correct with high probability but that it is not possible to verify
its correctness.  It generalizes to the case where the edge failure
probabilities are different, to computing the probability the graph is
not $k$-connected (for any fixed $k$), and to the more general problem
of approximating the {\em Tutte Polynomial} for a large family of
graphs.  It can also estimate the probability that an {\em Eulerian
  directed} graph remains strongly connected under edge failures.  Our
algorithm is easy to implement and appears likely to have satisfactory
time bounds in practice~\cite{Karger:ImpCut,Karger:ImpRel}.

Some care must be taken with the notion of approximation because
approximations are measured by {\em relative} error.  We therefore get
different results depending on whether we discuss the failure
probability $\fail(p)$ to the reliability (probability of remaining
connected) $\rel(p) =1-\fail(p)$.  Consider a graph with a very low
failure probability, say $\epsilon$.  In such a graph, approximating
$\rel(p)$ by $1$ gives a $(1+\epsilon)$-approximation to the
reliability, but approximating the failure probability by $0$ gives a
very poor (infinite) approximation ratio for $\fail(p)$.  Thus, the
failure probability is the harder quantity to approximate well.  On
the other hand, in a very unreliable graph, $\fail(p)$ becomes easy to
approximate (by 1) while $\rel(p)$ becomes the challenging quantity.
Our algorithm is an FPRAS for $\fail(p)$.  This means that in
extremely unreliable graphs, it cannot approximate $\rel(p)$.
However, it does solve the harder approximation problem on reliable
graphs, which are clearly the ones likely to be encountered in
practice.

The basic approach of our FPRAS is to consider two cases.  When
$\fail(p)$ is large, it can be estimated via direct Monte Carlo
simulation of random edge failures.  We thus focus on the case of
small $\fail(p)$.  Note that a graph becomes disconnected when all
edges in some cut fail (a {\em cut} is a partition of the vertices
into two groups; its edges are the ones with one endpoint in each
group).  The more edges cross a cut, the less likely it is that
they will all fail simultaneously.  We show that for small $\fail(p)$,
only the smallest graph cuts have any significant chance of failing.
We show that there is only a polynomial number of such cuts, and that
they can be enumerated in polynomial time.  We then use a {\em DNF
  counting} algorithm~\cite{Karp:DNF} to estimate the probability that
one of these explicitly enumerated cuts fails, and take this estimate
as an estimate of the overall graph failure probability.

After presenting our basic FPRAS for $\fail(p)$ in
Section~\ref{sec:reliability}, we present several extensions of it,
all relying on our observation regarding the number of small cuts a
graph can have.  In Section~\ref{sec:extensions}, we give FPRASs for
the network failure probability when every edge has a different
failure probability, for the probability that an Eulerian directed
graph fails to be strongly connected under random edge failures, and
for the probability that two particular ``weakly connected'' vertices
are disconnected by random edge failures.  In Section~\ref{sec:r-way},
we give an FPRAS for the probability that a graph partitions into more
than $r$ pieces for any fixed $r$.  In
Section~\ref{sec:deterministic}, we give two deterministic algorithms
for all-terminal reliability: a simple heuristic that provably gives
good approximations on certain inputs and a deterministic FPAS that
applies to a somewhat broader class of problems. In
Section~\ref{sec:Tutte}, we show that our techniques give an FPRAS for
the Tutte Polynomial on almost all graphs.

\subsection{Related Work}

Previous work gave algorithms for estimating $\fail(p)$ in certain
special cases.  Karp and Luby~\cite{Karp:Reliability} showed how to
estimate $\fail(p)$ in $n$-vertex planar graphs when the expected
number of edge failures is $O(\log n)$.  Alon, Frieze, and
Welsh~\cite{Alon:Tutte} showed how to estimate it when the input graph
is sufficiently dense (with minimum degree $\Omega(n)$).  Other
special case solutions are discussed in Colbourn's
survey~\cite{Colbourn}.  Lomonosov~\cite{Lomonosov:Mincut}
independently derived some of the results presented here.

A crucial step in our algorithm is the enumeration of minimum and
near-minimum cuts.  Dinitz et al.~\cite{Dinitz:Cactus} showed how to
enumerate (and represent) all minimum cuts.  Vazirani and
Yannakakis~\cite{Vazirani:EnumerateCuts} showed how to enumerate near
minimum cuts.  Karger and Stein~\cite{Karger:Contraction} gave faster
cut enumeration algorithms as well as bounds on the number of cuts
that we will use heavily.

A preliminary version of this work appeared
in~\cite{Karger:Reliability-Proc}.  The author's
thesis~\cite{Karger:Thesis} discusses reliability estimation in the
context of a general approach to random sampling in optimization
problems involving cuts.  In particular, this reliability work relies
on some new theorems bounding the number of small cuts in graphs;
these theorems have led to other results on applications of random
sampling to graph optimization
problems~\cite{Karger:Skeleton,Karger:Lincut,Karger:Stcut}.

\section{The Basic FPRAS}
\label{sec:reliability}

In this section, we present an FPRAS for $\fail(p)$.  
We use two methods, depending on the value of $\fail(p)$.

When $\fail(p)$ is large, we estimate it in polynomial time by direct
Monte Carlo simulation of edge failures. That is, we randomly fail
edges and check whether the graph remains connected.  Since $\fail(p)$
is large, a small number of simulations (roughly $1/\fail(p)$) gives
enough data to estimate it well.

When $\fail(p)$ is small, we resort to {\em cut enumeration} to
estimate it.  Observe that a graph becomes disconnected precisely when
all of the edges in some cut of the graph fail.  By a {\em cut} we
mean a partition of the graph vertices into two groups.  The {\em cut
edges} are those with one endpoint in each group (we also refer to
these edges as the ones {\em crossing} the cut).  The {\em value} of
the cut is the number of edges crossing the cut.

We show that when $\fail(p)$ is small, only small cuts of $G$ have any
significant chance of failing.  We observe that there is only a
polynomial number of such cuts that can be found in polynomial time.
We therefore estimate $\fail(p)$ by enumerating the polynomial-size
set of small cuts of $G$ and then estimating the probability that one
of them fails.

If each edge fails with probability $p$, then the probability that a
$k$-edge cut fails is $p^k$.  Thus, the smaller a cut, the more likely
it is to fail.  It is therefore natural to focus attention on the
small graph cuts.  Throughout this paper, we assume that our graph has
{\em minimum cut\/} value $c$---that is, that the smallest cut in the
graph has exactly $c$ edges.  Such a graph has a probability of at
least $p^c$ of becoming disconnected---namely, if the minimum cut
fails. That is:
\begin{fact}
If each edge of a graph with minimum cut $c$ fails independently with
probability $p$, then the probability that the graph becomes
disconnected is at least $p^c$.
\end{fact}

Clearly, the probability a cut fails decreases exponentially with the
number of edges in the cut.  This would suggest that a graph is most
likely to fail at its small cuts.  We formalize this intuition.

\begin{definition}
An {\em $\alpha$-minimum cut} is a cut with value at most $\alpha $
times the minimum cut value.
\end{definition}

Below, we show how to choose between the two approaches just
discussed.  If $\fail(p)\ge p^c \ge n^{-4}$ then, as we show in
Section~\ref{sec:Monte Carlo}, we can estimate it via Monte Carlo
simulation.  This works because $\Olog(1/\fail(p))=\Olog(n^4)$
experiments give us enough data to deduce a good estimate ($\Olog(f)$
denotes $O(f \log n)$). On the other hand, when $p^c < n^{-4}$, we
know that a given $\alpha$-minimum cut fails with probability
$p^{\alpha c} = n^{-4\alpha}$.  We show in Section~\ref{sec:countcuts}
that there are at most $n^{2\alpha}$ $\alpha$-minimum cuts.  It
follows that the probability that {\em any} $\alpha$-minimum cut fails
is less than $n^{-2\alpha}$---that is, exponentially decreasing with
$\alpha$.  Thus, for a relatively small $\alpha$, the probability that
a greater than $\alpha$-minimum cut fails is negligible.  Thus (as we
show in Section~\ref{sec:cut failure bounds}) we can 
approximate $\fail(p)$ by approximating the probability that some less
than $\alpha$-minimum cut fails.  Our FPRAS (in Section~\ref{sec:FPRAS
  small}) is based on enumerating these small cuts and determining the
probability that one of them fails.

\subsection{Monte Carlo Simulation}
\label{sec:Monte Carlo}

The most obvious way to estimate $\fail(p)$ is through Monte Carlo
simulations.  Given the failure probability $p$ for each edge, we can
``simulate'' edge failures by flipping an appropriately biased random
coin for each edge.  We can then test whether the resulting network is
connected.  If we do this many times, then the fraction of trials in
which the network becomes disconnected should intuitively provide a
good estimate of $\fail(p)$.  Karp and Luby~\cite{Karp:Reliability}
investigated this idea formally, and observed (a generalization of)
the following.

\begin{theorem}
  Performing $O((\log n)/(\epsilon^2\fail(p)))$ trials will give an
  estimate for $\fail(p)$ accurate to within $1\pm\epsilon$ with high
  probability.
\end{theorem}

\begin{corollary}
\labelprop{Corollary}{prop:large p}
If $\fail(p)\ge p^c \ge n^{-4}$, then $\fail(p)$ can be estimated to
within $(1+\epsilon)$ in $\Olog(mn^4/\epsilon^2)$ time using Monte Carlo
simulation.
\end{corollary}

The criterion that $\fail(p)$ not be too small can of course be
replaced by a condition that implies it.  For example, Alon, Frieze,
and Welsh~\cite{Alon:Tutte} showed that for any {\em constant} $p$,
there is an FPRAS for network reliability in {\em dense} graphs (those
with minimum degree $\Omega(n)$).  The reason is that as $n$ grows and
$p$ remains constant, $\fail(p)$ is bounded below by a constant on
dense graphs and can therefore be estimated in $\Olog(n^2/\epsilon^2)$
time by direct Monte Carlo simulation.

The flaw of the simulation approach is that it is too slow for small
values of $\fail(p)$, namely those less than 1 over a polynomial in
$n$.  It is upon this situation that we focus our attention for the
remainder of this section.  In this case, a huge number of standard
simulations would have to be run before we encountered a sufficiently
large number of failures to estimate $\fail(p)$ (note that we expect
to run $1/\fail(p)$ trials before seeing {\em any} failures).
Karp and Luby~\cite{Karp:Reliability} tackled this situation for
various problems, and showed that it could be handled in some cases by
biasing the simulation such that occurrences of the event being
estimated became more likely.  One of their results was an FPRAS for
network reliability in {\em planar} graphs, under the assumption that
the failure probability $p$ of edges is $O((\log n)/n)$ so that the
expected number of edges failing is $O(\log n)$.  Their algorithm is
more intricate than straightforward simulation, and, like ours, relies
on identifying a small collection of ``important cuts'' on which to
concentrate.

Another problem where direct Monte Carlo simulation breaks down, and
to which Karp and Luby~\cite{Karp:Reliability}, found a solution, is
that of {\em DNF counting:} given a boolean formula in disjunctive
normal form, and given for each variable a probability that it is set
to true, estimate the probability that the entire formula evaluates to
true.  Like estimating $\fail(p)$, this problem is hard when the
probability being estimated is very small. Karp and
Luby~\cite{Karp:Reliability} developed an FPRAS for DNF counting using
a biased Monte Carlo simulation.  The running time was later improved
by Karp, Luby, and Madras~\cite{Karp:DNF} to yield the following:

\begin{theorem}
There is an FPRAS for the DNF counting problem that runs in
$\Olog(s/\epsilon^2)$ time on a size $s$ formula.
\end{theorem}

We will use the DNF counting algorithm as a subroutine in our FPRAS.  

\subsection{Counting Near-minimum Cuts}
\label{sec:countcuts}

We now turn to the case of $p^c$ small.  We show that in this case,
only the smallest graph cuts have any significant chance of failure.
While it is obvious that cuts with fewer edges are more likely to
fail, one might think that there are so many large cuts that overall
they are more likely to fail than the small cuts.  However, the
following proposition lets us bound the number of large cuts and show
this is not the case.

\begin{theorem}
\labelprop{Theorem}{prop:countcuts}
 An undirected graph has less than $n^{2\alpha}$ $\alpha$-minimum
 cuts.
\end{theorem}

\begin{remark}
Vazirani and Yannakakis~\cite{Vazirani:EnumerateCuts} gave an
incomparable bound on the number of small cuts by {\em rank} rather
than by value.
\end{remark}

In this section, we sketch a proof of \ref{prop:countcuts}.  A
detailed proof of the theorem can be found
in~\cite{Karger:Contraction} and an alternative proof
in~\cite{Karger:Lincut}.  Here, we sketch enough detail to allow for
some of the extensions we will need later.  We prove the theorem only
for unweighted multigraphs (graphs with parallel edges between the
same endpoints); the theorem follows for weighted graphs if
we replace any weight $w$ edge by a set of $w$ unweighted parallel
edges.

\subsubsection{Contraction}

The proof of the theorem is based on the idea of {\em edge
  contraction}.  Given a graph $G=(V,W)$, and an edge $(v,w)$, we
  define a contracted graph $G/(v,w)$ with vertex set
  $V'=V\cup\set{u}-\set{v,w}$ for some new vertex $u$ and edge set
\[
E' = E-\set{(v,w)}\cup \set{(u,x) \mid (v,x) \in E \mbox{ or } (w,x) \in E}.
\]
In other words, in the contracted graph, vertices $v$ and $w$ are
replaced by a single vertex $u$, and all edges originally incident on
$v$ or $w$ are replaced by edges incident on $u$.  We also remove
self-loops formed by edges parallel to the contracted edge since they
cross no cut in the contracted graph.

\begin{fact}
\labelprop{Fact}{prop:cut correspondence}
There is a one-to-one correspondence between cuts in $G/e$ and cuts in
$G$ that $e$ does not cross.  Corresponding cuts have the same value.
\end{fact}
\begin{proof}
Consider a partition $(A,B)$ of the vertices of $G/(v,w)$.  The vertex $u$
corresponding to contracted edge $(v,w)$ is on one side or the other.
Replacing $u$ by $v$ and $w$ gives a partition of the vertices of
$G$.  The same edges cross the corresponding partitions.
\end{proof}

\subsubsection{The Contraction Algorithm}

We now use repeated edge contraction in an algorithm that selects a
cut from $G$.  Consider the following {\em Contraction Algorithm}.
While $G$ has more than $2$ vertices, choose an edge $e$ uniformly at
random and set $G \leftarrow G/e$.  When the algorithm terminates, we
are left with a two-vertex graph that has a unique cut.  A transitive
application of \ref{prop:cut correspondence} shows that this cut
corresponds to a unique cut in our original graph; we will say this
cut is {\em chosen} by the Contraction Algorithm.  We show that any
particular minimum cut is chosen with probability at least $n^{-2}$.
Since the choices of different cuts are disjoint events whose
probabilities add up to one, it will follow that there are at most
$n^2$ minimum cuts.  We then generalize this argument to
$\alpha$-minimum cuts.

\begin{lemma}
The Contraction Algorithm chooses any particular minimum cut with
probability at least $n^{-2}$.
\end{lemma}
\begin{proof}
  Each time we contract an edge, we reduce the number of vertices in
  the graph by one.  Consider the stage in which the graph has $r$
  vertices. Suppose $G$ has minimum cut $c$.  It must have minimum
  degree $c$, and thus at least $rc/2$ edges.  Our particular minimum
  cut has $c$ edges.  Thus a randomly chosen edge is in the minimum
  cut with probability at most $c/(rc/2) = 2/r$.  The probability that
  we never contract a minimum cut edge through all $n-2$ contractions
  is thus at least
\begin{eqnarray*}
\left(1-\frac{2}{n}\right)\left(1-\frac{2}{n-1}\right)\cdots\left(1-\frac{2}{3}\right)
         &=
         &\left(\frac{n-2}{n}\right)\left(\frac{n-3}{n-1}\right)\cdots\left(\frac{2}{4}\right)\left(\frac{1}{3}\right)\\
         &= &\frac{\quad\quad\quad\quad\quad (n-2)(n-3)\cdots
         (3)(2)(1)}{n(n-1)(n-2)\cdots\cdots(4)(3)\quad} \\
         &= &\frac{2}{n(n-1)}\\
         &= &\binom{n}{2}^{-1}\\
        &> &n^{-2}.
\end{eqnarray*}
\end{proof}

\subsubsection{Proof of \protect{\ref{prop:countcuts}}}

We can extend the approach above to prove \ref{prop:countcuts}.  We
slightly modify the Contraction Algorithm and lower bound the
probability it chooses a particular $\alpha$-minimum cut.  With $r$
vertices remaining, the probability we choose an edge from our
particular $\alpha$-minimum cut is at most $2\alpha/r$.  Let
$k=\ceil{2\alpha}$.  Suppose we perform random contractions until we
have a $k$-vertex graph.  In this graph, choose a vertex partition
uniformly at random, so that each of its cuts is chosen with
probability $2^{1-k}$.  It follows that a particular $\alpha$-minimum
cut is chosen with probability
\begin{eqnarray*}
(1-\frac{2\alpha}{n})(1-\frac{2\alpha}{(n-1)})\cdots(1-\frac{2\alpha}{k+1}) 2^{1-k}
        &= &\frac{(n-2\alpha)!}{(k-2\alpha)!}\frac{k!}{n!}2^{1-k}\\
        &= &\frac{\binom{k}{2\alpha}}{\binom{n}{2\alpha}}2^{1-k}\\
        &> &n^{-2\alpha}.
\end{eqnarray*}
Note that for $\alpha$ not a half-integer, we are making use of {\em
  generalized binomial coefficients} which may have non-integral
arguments.  These are discussed
in~\cite[Sections~1.2.5--6]{Knuth:Book1} (cf. Exercise 1.2.6.45).
There, the Gamma function is introduced to extend factorials to real
numbers such that $\alpha! = \alpha(\alpha-1)!$ for all real
$\alpha>0$.  Many standard binomial identities extend to generalized
binomial coefficients, including the facts that $\binom{n}{2\alpha} <
n^{2\alpha}/(2\alpha)!$ and $2^{2\alpha-1} \le (2\alpha)!$ for $\alpha
\ge 1$.

\begin{remark}
The Contraction Algorithm described above is used only to count
cuts.  An efficient {\em implementation} given
in~\cite{Karger:Contraction} can be used to {\em find} all
$\alpha$-minimum cuts in $\Olog(n^{2\alpha})$ time.  We  use this
algorithm in our FPRAS.
\end{remark}

\subsection{Cut Failure Bounds}
\label{sec:cut failure bounds}

Using the cut counting theorem just given, we show that large cuts do
not contribute significantly to a graph's failure probability.
Consider \ref{prop:countcuts}.  Taking $\alpha=1$, it follows from the
union bound that the probability that some minimum cut fails is at
most $n^2p^c$.  We now show that the probability that {\em any} cut
fails is only a little bit larger.

\begin{theorem}\labelprop{Theorem}{prop:reliable}
Suppose a graph has minimum cut $c$ and that each edge of the graph
fails independently with probability $p$, where $p^c =
n^{-(2+\delta)}$ for some $\delta>0$.  Then
\begin{enumerate}
\item\label{failure} The probability that the given graph 
disconnects is at most $n^{-\delta}(1+2/\delta)$, and
\item\label{large failure} The probability that a cut of value $\alpha
  c$ or greater fails in the graph is at most $n^{-\alpha\delta
    }(1+2/\delta)$.
\end{enumerate}
\end{theorem}
\begin{remark}
We conjecture that a probability bound of $n^{-\alpha\delta}$ can be
proven (eliminating the $(1+2/\delta)$ term).
\end{remark}
\begin{proof}
  We prove Part~\ref{failure} and then note the small change needed to
  prove Part~\ref{large failure}.  For the graph to become
  disconnected, all the edges in some cut must fail.  We therefore
  bound the failure probability by summing the probabilities that each
  cut fails.  Let $r$ be the number of cuts in the graph, and let
  $c_1,\ldots,c_{r}$ be the values of the $r$ cuts in increasing order
  so that $c = c_1 \le c_2\le\cdots\le c_{r}$.  Let $p_k=p^{c_k}$ be
  the probability that all edges in the $k^{th}$ cut fail.  Then the
  probability that the graph disconnects is at most $\sum p_k$, which
  we proceed to bound from above.

We proceed in two steps.  First, consider the first $n^2$ cuts in the
ordering (they might not be minimum cuts).  Each of them has $c_k \ge
c$ and thus has $p_k \le n^{-(2+\delta)}$, so that
\[
\sum_{k \le n^2} p_k \le (n^2)(n^{-(2+\delta)}) = n^{-\delta}.
\]
Next, consider the remaining larger cuts.  According to
\ref{prop:countcuts}, there are less than $n^{2\alpha}$ cuts of
value at most $\alpha c$.  Since we have numbered the cuts in
increasing order, this means that $c_{n^{2\alpha}} > \alpha c$. In
other words, writing $k=n^{2\alpha}$,
\[
c_k > \frac{\ln k}{2 \ln n} \cdot c
\]
and thus 
\begin{eqnarray*}
p_k &< &(p^c)^{\frac{\ln k}{2 \ln n}}\\
 &= &(n^{-(2+\delta)})^{\frac{\ln k}{2 \ln n}}\\
 &= &k^{-(1+\delta/2)}.
\end{eqnarray*}
It follows that 
\begin{eqnarray*}
\sum_{k > n^2} p_k &< &\sum_{k>n^2} k^{-(1+\delta/2)}\\
                &\le &\int_{n^2}^\infty k^{-(1+\delta/2)}\,dk\\
                &\le &2n^{-\delta}/\delta.
\end{eqnarray*}
Summing the bounds for the first $n^2$ and for the remaining cuts
gives a total of $n^{-\delta}+2n^{-\delta}/\delta$, as claimed.

The proof of Part~\ref{large failure} is the same, except that we sum
only over those cuts of value at least $\alpha c$.
\end{proof}

\begin{remark}
  A slightly stronger version of Part~\ref{failure} was first proved
  by Lomonosov and Polesskii~\cite{Lomonosov:Reliability} using
  different techniques that identified the cycle as the most
  unreliable graph for a given $c$ and $n$.  We sketch their result,
  which we need for a different purpose, in
  Section~\ref{sec:Lomonosov}.  However, Part~\ref{large failure} is
  necessary for the FPRAS and was not previously known.
\end{remark}

\subsection{An Approximation Algorithm}
\label{sec:FPRAS small}

Our proof that only small cuts matter leads immediately to an FPRAS.
First we outline our solution.  Given that $\fail(p) < n^{-4}$,
\ref{prop:reliable} shows that the probability that a cut of value
much larger than $c$ fails is negligible, so we need only
determine the probability that a cut of value near $c$ fails.  We do
this as follows.  First, we enumerate the (polynomial size) set of
near-minimum cuts that matter.  From this set we generate a polynomial
size boolean expression (with a variable for each edge, true if the
edge has failed) that is true if any of our near-minimum cuts has
failed.  We then need to determine the probability that this boolean
expression is true; this can be done using the DNF counting techniques
of Karp, Luby, and Madras~\cite{Karp:Reliability,Karp:DNF}.  Details
are given in the following theorem.

\begin{theorem}
\labelprop{Theorem}{prop:small p}
  When $\fail(p)<n^{-4}$, there is a (Monte Carlo) FPRAS for
  estimating $\fail(p)$ running in $\Olog(mn^4/\epsilon^3)$ time.
\end{theorem}
\begin{proof}
  Under the assumption, the probability that a particular minimum cut
  fails is $p^c \le \fail(p) \le n^{-4}$. We show there is a constant
  $\alpha$ for which the probability that any cut of value greater
  than $\alpha c$ fails is at most $\epsilon\fail(p)$.
  This proves that to approximate to the desired accuracy we need only
  determine the probability that some cut of value less than $\alpha
  c$ fails.  It remains to determine $\alpha$.  Write
  $p^c=n^{-(2+\delta)}$; by hypothesis $\delta\ge 2$.  Thus by
  \refprop{prop:reliable}, the probability that a cut larger than
  $\alpha c$ fails is at most $2n^{-\delta\alpha}$.  On the other
  hand, we know that $n^{-(2+\delta)}=p^c \le \fail(p)$, so it suffices to
  find an $\alpha$ for which $2n^{-\delta\alpha} \le \epsilon
  n^{-(2+\delta)}$.  Solving shows that $\alpha
  =1+2/\delta-(\ln(\epsilon/2))/\delta \ln n \le
  2-\ln(\epsilon/2)/2\ln n$ suffices and that we therefore need only
  examine the smallest $n^{2\alpha} = O(n^{4}/\epsilon)$ cuts.

  We can enumerate these cuts in $O(n^{2\alpha}\log^3 n)$ time using
  certain randomized algorithms~\cite{Karger:Fastcut,Karger:Lincut} (a
  somewhat slower deterministic algorithm can be found
  in~\cite{Vazirani:EnumerateCuts}). Suppose we assign a boolean
  variable $x_e$ to each edge $e$; $x_e$ is true if edge $e$ fails and
  false otherwise.  Therefore, the $x_e$ are independent and true with
  probability $p$. Let $E_i$ be the set of edges in the $i^{th}$ small
  cut. Since the $i^{th}$ cut fails if and only if all edges in it
  fail, the event of the $i^{th}$ small cut failing can be written as
  $F_i = \wedge_{e \in E_i}x_e$.  Then the event of some small cut
  failing can be written as $F = \vee_i F_i$.  We wish to know the
  probability that $F$ is true.  Note that $F$ is a formula in
  disjunctive normal form.  The size of the formula is equal to the
  number of clauses $(n^{2\alpha})$ times the number of variables per
  clause (at most $\alpha c$), namely $O(cn^{2\alpha})$.  The FPRAS of
  Karp, Luby, and Madras~\cite{Karp:DNF} estimates the truth
  probability of this formula, and thus the failure probability of the
  small cuts, to within $(1\pm\epsilon)$ in
  $\Olog(cn^{2\alpha}/\epsilon^2) =
  \Olog(cn^4/\epsilon^{3})=\Olog(mn^4/\epsilon^3)$ time.

We are therefore able to estimate to within $(1\pm\epsilon)$ the value
of a probability (the probability that some $\alpha$-minimum cut
fails) that is within $(1\pm\epsilon)$ of the probability of the event
we really care about (the probability that some cut fails).  This
gives us an overall estimate accurate to within $(1\pm\epsilon)^2
\approx (1\pm2\epsilon)$.
\end{proof}

\begin{corollary}
There is an FPRAS for $\fail(p)$ running in $\Olog(mn^4/\epsilon^3)$
time.
\end{corollary}
\begin{proof}
  Suppose we wish to estimate the failure probability to within a
  $(1\pm\epsilon)$ ratio.  If $\fail(p) > n^{-4}$, then we estimate it
  in $\Olog(mn^4/\epsilon^2)$ time by direct Monte Carlo simulation as
  in \ref{prop:large p}.  Otherwise, we can run the
  $\Olog(mn^4/\epsilon^3)$ time algorithm of \ref{prop:small p}.
\end{proof}

If the graph is sparse (with $O(n)$ edges) and the minimum cut is
$\Olog(1)$ (both these conditions apply to, e.g., planar graphs) then
the time for a Monte Carlo trial is $O(n)$, while the size of the
formula for the DNF counting step above is $\Olog(n^{2\alpha})$. So if
we use a different $\fail(p)$ threshold for deciding which algorithm
to use, we can improve the running time bound to
$\Olog(n^{3.8}/\epsilon^2)$.

While this time bound is still rather poor, experiments have suggested
that performance in practice is significantly better---typically
$\Olog(n^3)$ on sparse graphs~\cite{Karger:ImpRel}.

\section{Extensions}
\label{sec:extensions}

We now discuss several extensions of our basic FPRAS.  In this
section, we will consider many cases in which it is sufficient to
consider the probability that an $\alpha$-minimum cut fails for some
$\alpha=O(1-\log \epsilon / \log n)$ (as in the previous section) that
is understood in context but not worth deriving explicitly.  We will
refer to these $\alpha$-minimum cuts as the {\em weak} cuts of the
graph.

\subsection{Varying Failure Probabilities}

The analysis and algorithm given above extend to the case where each
edge $e$ has its own failure probability $p_e$.  To extend the
analysis, we transform a graph with varying edge failure probabilities
into one with identical failure probabilities.  Given the graph $G$
with specified edge failure probabilities, we build a new graph $H$
all of whose edges have the same failure probability $p$, but that has
the same failure probability as $G$.  Choose a small parameter
$\theta$.  Replace an edge $e$ of failure probability $p_e$ by a
``bundle'' of $k_e$ parallel edges, each with the same endpoints as
$e$ but with failure probability $1-\theta$, where
\[
k_e = \ceil{-(\ln p_e)/\theta}.
\]
This bundle of edges keeps its endpoints connected unless all the
edges in the bundle fail; this happens with probability
\[
(1-\theta)^{\ceil{-(\ln p_e)/\theta}}.
\]
As $\theta\rightarrow 0$, this failure probability converges to $p_e$.
Therefore, the reliability of $H$ converges as $\theta \rightarrow 0$
to the reliability of $G$.   Thus, to determine the failure
probability of $G$, we need only determine the failure probability of
$H$ in the limit as $\theta \rightarrow 0$.

Since $H$ has all edge failure probabilities the same, our
Section~\ref{sec:reliability} analysis of network reliability applies
to $H$.  In particular, we know that it suffices to enumerate the weak
cuts of $H$ and then determine the probability that one of them fails.
To implement this idea, note that changing the parameter $\theta$
scales the values of cuts in $H$ without changing their relative
values (modulo a negligible rounding error).  We therefore build a
weighted graph $F$ by taking graph $G$ and giving a weight $(\ln
1/p_e)$ to edge $e$.  The weak cuts in $F$ correspond to the weak cuts
in $H$.  We find these weak cuts in $F$ using the Contraction
Algorithm (which works for weighted graphs~\cite{Karger:Contraction})
as before.

Given the weak cuts in $H$, we need to determine the limiting probability that
one of them fails as $\theta \rightarrow 0$.  We have already argued
that as $\theta \rightarrow 0$, the probability a cut in $H$ fails
converges to the probability that the corresponding cut in $G$ fails.
Thus we actually want to determine the probability that one of a given
set of cuts in $G$ fails.  We do this as before: we build a boolean
formula with variables for the edges of $G$ and with a clause for each
weak cut that is true if all the edges of the cut fail.  The only
change is that variable $x_e$ is set to true with probability $p_e$.
The algorithm of~\cite{Karp:DNF} works with these varying truth
probabilities and computes the desired quantity.  This gives:

\begin{theorem}
There is an FPRAS for the all terminal network reliability problem
with varying edge failure probabilities.
\end{theorem}

One might be concerned by the use of logarithms to compute edge
weights.  However, it is easy to see that in fact approximate
logarithms suffice for the purpose of enumerating small cuts.  If we
approximate each logarithm to within relative error $.1$, then every
$\alpha$-minimum cut in $F$ remains an $11\alpha/9$-minimum cut in the
approximation to $F$.  Thus we can enumerate a slightly larger set of
near-minimum cuts in order to find the weak cuts.  Once we find the
weak cuts, we use the original $p_e$ values in the DNF counting
algorithm.

In the case of varying failure probabilities, we cannot bound the
number of edges in any particular weak cut by a quantity less than $m$
(a weak cut may have $m-n$ edges with large failure probabilities).
Thus the size of the DNF formula, and thus the running time of the DNF
counting algorithm, may be as large as $mn^{2\alpha}\approx mn^4/\epsilon$.

All the other extensions described in this paper can also be modified
to handle varying failure probabilities.  But for simplicity, we focus
on the uniform case.

\subsection{Multiterminal Reliability}

The multiterminal reliability problem is a generalization of the
all-terminal reliability problem.  Instead of asking whether the graph
becomes disconnected, we consider a subset $K$ of the vertices and ask
if some pair of them becomes disconnected.  If some pair of vertices
in $K$ is separated by a cut of value $O(c)$, then we can use the same
theorem on the exponential decay of cut failure probabilities to prove
that we only need to examine the small cuts in the graph to determine
whether some pair of vertices in $K$ becomes disconnected.

\begin{lemma}
If some pair of vertices in $K$ is separated by a cut of value $O(c)$,
then there is an FPRAS for the multiterminal reliability problem with
source vertices $K$.
\end{lemma}
\begin{proof}
We focus on the case of uniform failure probability $p$; the
generalization to arbitrary failure probabilities is as before.
Suppose a cut of value $\beta c$ separates vertices in $K$.  Then the
probability that $K$ gets disconnected when edges fail with
probability $p$ is at least $p^{\beta c}$.  If $p^{c} > n^{-4}$, then
$p^{\beta c} > n^{-4\beta}=n^{-O(1)}$ and we use Monte Carlo
simulation as before to estimate the failure probability.  If $p^{c} <
n^{-4}$, then by \ref{prop:reliable}, the probability that a cut of
value exceeding $\alpha c$ fails is $O(n^{-2\alpha})$.  Thus, choosing
$\alpha$ such that $n^{-2\alpha}\le\epsilon p^{\beta c}$, we can
enumerate the weak cuts and apply DNF counting.
\end{proof}

\subsection{$k$-Connectivity}

Just as we estimated the probability that the graph fails to be
connected, we can estimate the probability that it fails to be $k$-edge
connected for any constant $k$.  Note that the graph fails to be
$k$-edge connected only if some cut has less than $k$ of its edges
survive.  The probability of this event decays exponentially with the
value of the cut, allowing us to prove (as with
\refprop{prop:reliable}) that if the probability that fewer than $k$
edges in a minimum cut survive is $O(n^{-(2+\delta)})$, then the
probability that fewer than $k$ edges survive in a non-weak cut is
negligible. Thus, if direct Monte Carlo simulation is not applicable,
we need only determine the probability that some weak cut keeps less
than $k$ of its edges.  But this is another DNF counting problem.  For
any particular weak cut containing $C\le m$ edges, we enumerate all
$\binom{C}{C-k+1} = O(C^{k-1}) = O(m^{k-1})$ sets of $C-k+1$ edges,
and for each add a DNF clause that is true if all the given edges
fail.

In fact, one can also adapt the algorithm of~\cite{Karp:DNF} to
determine the probability that all but $k-1$ variables in some clause of
a DNF formula become true; thus we can continue to work with the
$O(mn^4/\epsilon)$ size formula we used before.

\begin{corollary}
For any constant $k$, there is an FPRAS for the probability that a graph
with edge failure probabilities fails to be $k$-edge connected.
\end{corollary}

\subsection{Eulerian Directed Graphs}

A natural generalization of the all-terminal reliability problem to
directed graphs is to ask for the probability that a directed graph
with random edge failures remains strongly connected.  A directed
graph fails to be strongly connected precisely when all the edges in
some {\em directed} cut fail.  In general, the techniques of this
paper cannot be applied to directed graphs---the main reason being
that a directed graph can have exponentially many minimum directed
cuts.

We can, however, handle one special case.  In an {\em Eulerian}
directed graph $G$ on vertex set $V$, the number of edges crossing
from any vertex set $A$ to $V-A$ is equal to the number of edges
crossing from $V-A$ to $A$.  Thus if we construct an undirected graph
$H$ by removing the directions from the edges of $G$, we know that any
(directed) cut in $G$ has value equal to half that of the
corresponding (undirected) cut in $H$.  It follows that the
$\alpha$-minimum directed cuts of $G$ correspond to $\alpha$-minimum
undirected cuts of $H$.  Therefore, there are at most $2n^{2\alpha}$
$\alpha$-minimum directed cuts in $G$ that can be enumerated by
enumerating the $\alpha$-minimum cuts of $H$ (the factor of 2 arises
from considering both directions for each cut).  As in the undirected
case, if the directed failure probability is less than $n^{-4}$, an
analogue of \ref{prop:reliable} immediately follows, showing that only
weak directed cuts are likely to fail. It therefore suffices to
enumerate a polynomial number of weak directed cuts to estimate the
directed failure probability.

\begin{corollary}
There is an FPRAS for the probability that a directed Eulerian graph
fails to remain strongly connected under random edges failures.
\end{corollary}

\begin{corollary}
  For any constant $k$ there is an FPRAS for the probability that a
  directed Eulerian graph fails to have directed connectivity $k$
  under random edges failures.
\end{corollary}

\subsection{Random Orientations}
\label{sec:orientations}

In a similar fashion, we can estimate the probability that, if we
orient each edge of the graph randomly, the graph fails to be strongly
connected.  For each cut, we make a DNF formula with two clauses, one
of which is true if all edges point ``left'' and the other if all
edges point ``right.''  (This observation is due to Alan Frieze.)
This problem can also be phrased as estimating the number of
non-strongly connected orientations of an undirected graph; in this
form, it is related to the {\em Tutte Polynomial} discussed in
Section~\ref{sec:Tutte}.  Similarly, we can estimate the probability
that random orientations fail to produce a $k$-connected directed
graph.

\section{Partition into $r$ Components}
\label{sec:r-way}

The quantity $\fail(p)$ is an estimate of the probability that the
graph partitions into more than one connected component.  We can
similarly estimate the probability that the graph partitions into $r$
or more components for any constant $r$.  Besides its intrinsic
interest, the analysis of this problem will be important in our study
of some heuristics and derandomizations in
Section~\ref{sec:deterministic} and the Tutte Polynomial in
Section~\ref{sec:Tutte}.

We first note that a graph partitions into $r$ or more components only
if an {\em $r$-way cut}---the set of edges with endpoints in different
components of an $r$-way vertex partition---loses all its edges.  Note
that some of the vertex sets of the partition might induce disconnected
subgraphs, so that the $r$-way partition might induce more than $r$
connected components.  However, it certainly does not induce less.
Our approach to $r$-way reliability is the same as for the $2$-way
case: we show that there are few small $r$-way cuts and that
estimating the probability one fails suffices to approximate the
$r$-way failure probability.  As a corollary, we show that the
probability of $r$-way partition is much less than that of $2$-way
partition.

\subsection{Counting Multiway Cuts}

We enumerate multiway cuts using the Contraction Algorithm as for
$2$-way case.  Details can be found in~\cite{Karger:Contraction}.

\begin{lemma}
In an $m$-edge unweighted graph the minimum $r$ way cut has value at
most $2m(r-1)/n$.
\end{lemma}
\begin{proof}
  A graph's average degree is $2m/n$.  Consider an $r$-way cut with
  each of the $r-1$ vertices of smallest degree as its own singleton
  component and all the remaining vertices as the last component.  The
  value of this cut is at most the sum of the singleton vertex
  degrees, which is at most $r-1$ times the average degree.
\end{proof}

\begin{corollary}
There are at most $\binom{n}{2(r-1)}$ minimum $r$-way cuts.
\end{corollary}
\begin{proof}
Fix a particular $r$-way minimum cut and run the Contraction Algorithm
until we have $2(r-1)$ vertices.  By the previous lemma, the probability
that we pick an edge of our fixed cut when $k$ vertices remain is at most
$2\frac{r-1}{k}$.  Thus the probability that our fixed minimum
$r$-way cut is chosen is
\[
\prod_{k=2r-1}^n \left (1-\frac{2(r-1)}{k}\right)
\]
which is analyzed exactly as in the proof of \ref{prop:countcuts},
substituting $r-1$ for $\alpha$.
\end{proof}

\begin{corollary}
\labelprop{Corollary}{prop:count r-way}
For arbitrary $\alpha \ge 1$, there are at most $(rn)^{2\alpha(r-1)}$
$\alpha$-minimum $r$-way cuts that can be enumerated in
$\Olog((rn)^{2\alpha(r-1)})$ time.
\end{corollary}
\begin{proof}
First run the Contraction Algorithm until the number of vertices
  remaining is $\ceil{2\alpha(r-1)}$.  At this point, choose a random
  $r$-way partition of what remains.  There are at most
  $r^{2\alpha(r-1)}$ such partitions.

  The time bound follows from the analysis of the Recursive Contraction
  Algorithm~\cite{Karger:Contraction}.
\end{proof}

\begin{remark}
  We conjecture that in fact the correct bound is $O(n^{\alpha r})$
  $\alpha$-minimum $r$-way cuts.  Section~\ref{sec:Lomonosov} shows
  this is true for $\alpha=1$.  Proving it for general $\alpha$ would
  slightly improve our exponents in the following sections.
\end{remark}

\subsection{An Approximation Algorithm}

Our enumeration of multiway cuts allows an analysis and reduction to
DNF counting exactly analogous to the one performed for
$\fail(p)$.

\begin{corollary}
Suppose a graph has $r$-way minimum cut value $c_r$ and that each edge
fails with probability $p$, where $p^{c_r} = (rn)^{-(2+\delta)(r-1)}$
for some constant $\delta > 0$.  Then the probability that an
$\alpha$-minimum $r$-way cut fails is at most
$(rn)^{-\alpha\delta(r-1)}(1+2/\delta)$
\end{corollary}
\begin{proof}
Exactly as for \ref{prop:reliable}, substituting $(rn)^{(r-1)}$ (drawn
from \ref{prop:count r-way}) for $n$ everywhere.
\end{proof}

\begin{corollary}
\labelprop{Corollary}{prop:multiway}
There is an algorithm for $\epsilon$-approximating the probability
that a graph partitions into $r$ or more components, running in
$\Olog(m(rn)^{4(r-1)}/\epsilon^3)$ time.
The algorithm is an FPRAS with running time
$\Olog(mn^{4(r-1)}/\epsilon^3)$ for any fixed $r$.
\end{corollary}
\begin{proof}
  Exactly as for the two-way cut case, with $(rn)^{(r-1)}$ replacing
  $n$ everywhere.  Let $c_r$ be the $r$-way minimum cut value and let
  $\delta$ be defined by $p^{c_r} = (rn)^{-(2+\delta)(r-1)}$.  If $p^{c_r} >
  (rn)^{-4(r-1)}$, estimate the partition probability via Monte
  Carlo simulation. Otherwise, it follows as in the $2$-way cut case
  that for the same constant $\alpha$ as we chose there, the
  probability that a greater than $\alpha$-minimum $r$-way cut fails
  is less than $\epsilon p^{c_r}$.  Thus to estimate the partition
  probability it suffices to enumerate (in
  $\Olog((rn)^{4(r-1)}/\epsilon)$ time) the set of $\alpha$-minimum
  cuts and perform DNF counting.  
\end{proof}

One might wish to compute the probability that a graph partitions into
{\em exactly} $r$ components, but it is not clear that this can be
done.  In particular, computing $\rel(p)$ can be reduced to this
problem (for any $r$) by adding $r-1$ isolated vertices.  There is
presently no known FPRAS for $\rel(p)$.

\subsection{Comparison to 2-way Cuts}

For Sections~\ref{sec:deterministic} and~\ref{sec:Tutte}, we need to
show that the probability of partition into $r$ components is much
less than that of partition into $2$ components.  We give two proofs,
the first simpler but with a slightly weaker bound.  The following
sections can use the weaker analysis at the cost of worse exponents.
In this section, the term ``cut'' refers exclusively to two-way cuts
unless we explicitly modify it.

\subsubsection{A simple argument}

\begin{lemma}
\labelprop{Lemma}{prop:r-way vs 2-way simple}
If $p^c = n^{-(2+\delta)}$, then the probability that an $r$-way cut
fails is at most $n^{-\delta r/4}(1+2/\delta)$.
\end{lemma}
\begin{proof}
  We show that any $r$-way cut contains the edges of a (2-way) cut of
  value $rc/4$.  Thus, if an $r$-way cut fails then an
  $(r/4)$-minimum cut fails.  The probability that this happens has
  been upper-bounded by \ref{prop:reliable}.

  To show the claim, consider an $r$-way cut.  Contract each component
  of the $r$-way partition to a single vertex, yielding an $r$-vertex
  graph $G'$.  All edges in this graph correspond to edges of the
  $r$-way cut.  Every cut in $G'$ corresponds to a cut of the same
  value in the original graph, so it suffices to show that $G'$ has a
  $2$-way cut of value $rc/4$.  To see this, note that every vertex in
  $G'$ has degree at least $c$, so the number of edges in $G'$ is at
  least $rc/2$.  Consider a random cut of $G'$, generated by assigning
  each vertex randomly to one side or the other.  Each edge has a
  $1/2$ chance of being cut by this partition, so the expected value
  of this cut is at least $rc/4$.  It follows that $G'$ has a cut of
  value at least $rc/4$ that corresponds to a cut of value at least
  $rc/4$ in the original graph.
\end{proof}

\subsubsection{A better argument}
\label{sec:Lomonosov}

We can get a slightly better bound on the probability that a graph
partitions into $r$ components via a small variation on an argument
due to Lomonosov and
Polesskii~\cite{Lomonosov:Reliability,Lomonosov:MatroidSampling,Colbourn}.
The better bound improves some of our exponents.  Their proof uses
techniques somewhat different from the remainder of the paper and can
safely be skipped.

\begin{lemma}
\labelprop{Lemma}{prop:r-way vs cycle}
Let $\fail_r(G,p)$ denote the probability that $G$ partitions into $r$
or more connected components when each edge fails with probability
$p$.  Let $G$ have minimum cut $c$ for some even $c$.  Let $C_n$ be a
cycle with $c/2$ edges between adjacent vertices.  Then for any $r$,
$\fail_r(G,p) \le \fail_r(C_n,p)$.
\end{lemma}

\begin{corollary}
\labelprop{Corollary}{prop:r-way vs 2-way}
\labelprop{Corollary}{prop:r-way failure} For any graph $G$ with
minimum cut $c$, if edges fail with probability $p$ where
$p^c=n^{-(2+\delta)}$, then the probability the failed graph has $r$ or
more connected components is less than $n^{-\delta r
  /2}$.
\end{corollary}

\begin{remark} 
  Note that for $r=2$, the above result gives a slightly stronger
  bound on $\fail(p)$ than we are able to get in \ref{prop:reliable}.
  Unfortunately, this argument does not appear to extend to proving
  the bound we need on the probability that a greater than
  $\alpha$-minimum $r$-way cut fails.
\end{remark}

\begin{proofof}{\ref{prop:r-way failure}}
Thanks to \ref{prop:r-way vs cycle}, it suffices to prove this claim
for the case of $G$ a cycle $C_n$ with $(c/2)$-edge ``bundles''
between adjacent vertices.  The number of components into which $C_n$
is partitioned is equal to the number of bundles which fail, so we
need only bound the probability that $r$ or more bundles fail.  The
probability that a single bundle fails is $p^{c/2} =
n^{-(1+\delta/2)}$, so the probability $r$ particular bundles fail is
$n^{-r(1+\delta/2)}$.  There are $\binom{n}{r}<n^r$ sets of exactly
$r$ bundles.  It follows that the probability $r$ or more bundles fail
is less than $n^r n^{-r(1+\delta/2)}=n^{-r\delta/2}$.
\end{proofof}

\begin{proofof}{\ref{prop:r-way vs cycle}}
Consider the following time-evolving version of the Contraction
Algorithm on a connected graph $G$.  Each edge of $G$ is given an {\em
arrival time} chosen independently from the exponential distribution
with mean $1$.  Each time an edge arrives, we contract its endpoints
if they have not already been contracted.  This gives rise to a
sequence of graphs $G=G_n, G_{n-1}, \ldots, G_1$ where $G_r$ has $r$
vertices.  Let $G[t]$ be the graph that exists at time $t$.  Thus
initially $G[0]=G_n$ and eventually $G[\infty]$ has one vertex since
all edges have arrived.  We draw a correspondence between this model
and our edge failure model as follows: at time $t$, the failed edges
are those which have not yet arrived.  It follows that each vertex in
$G[t]$ corresponds to a connected component of $G$ when each edge has
failed (to arrive) independently with probability $e^{-t}$.

We consider the random variable $T_r(G)$ defined as the time at which
the edge that contracts $G_r$ to $G_{r-1}$ arrives.  We show that
$T_r(C_n)$ {\em stochastically dominates} $T_r(G)$ for every
$r$---that is,
\[
\Pr[T_r(G) \ge t] \le \Pr[T_r(C_n) \ge t]
\]
(See Motwani and Raghavan~\cite{Motwani:RandomizedAlgorithms} for
additional discussion of this definition).  Assuming this is true, we
can prove our result as follows:
\begin{eqnarray*}
\Pr[\mbox{$G[t]$ has $r$ or fewer components}] 
&= &\Pr[T_r(G) \le t]\\
&\ge &\Pr[T_r(C_n) \le t]\\
&= &\Pr[\mbox{$C_n[t]$ has $r$ or fewer components}] 
\end{eqnarray*}

To prove stochastic domination, let $t_r(G)=T_{r-1}(G)-T_{r}(G)$
denote the length of time for which $G_r$ exists before being
contracted to $G_{r-1}$.  Clearly, $t_r(G)$ is just the time it
takes for an edge to arrive that has endpoints in different connected
components of $G_{r}$.  Thanks to the memoryless nature of the
exponential distribution, the $t_r$ are mutually independent (this
will be justified more carefully later).  It follows that $T_r(G) =
\sum_{r'=r}^n t_{r'}(G)$.  Similarly, $T_r(C_n) = \sum_{r'=r}^n
t_{r'}(C_n)$.  We will show that $t_r(C_n)$ stochastically dominates
$t_r(G)$ for every $r$.  The fact that $T_r(C_n)$ stochastically
dominates $T_r(G)$ then follows from the fact that when $X$ dominates
$X'$ and $Y$ dominates $Y'$ and the variables are independent, $X+Y$
dominates $X'+Y'$.

To analyze $t_r$, suppose there are $m_{r}$ edges in $G_r$ (note
$m_{r}$ is a random variable).  The arrival time of each edge in $G_r$
measured from $T_r(G)$ is exponentially distributed with mean 1.
Therefore, the arrival time of the first such edge, namely $t_r(G)$,
is exponentially distributed with mean $1/m_{r}$.  Now note that $G_r$
is $c$-connected, so it must have $m_{r} \ge cr/2$.  It follows that
$t_{r}(G)$ is exponentially distributed with mean at most $2/cr$,
meaning that it is stochastically dominated by any exponentially
distributed variable with mean $2/cr$.  On the other hand, when $C_n$
has been reduced to $r$ components, it is isomorphic to $C_{r}$.  By
the same analysis as for $G$, we know $t_r(C_n)$ is exponentially
distributed with mean $2/cr$, and thus stochastically dominates
$t_r(G)$.

Our glib claim that the $t_r$ are independent needs some additional
justification.  Technically, we condition on the values
$G_n,\ldots,G_1$ of the evolving graph.  We show that regardless of
what values we condition on, $T_r(C_n)$ stochastically dominates
$T_r(G \mid G_n,\ldots,G_1)$.  Since the stochastic domination applies
regardless of our conditioning event, it follows even if we do not
condition.

Once we have conditioned on the value $G_r$, $t_r$ is just the time it
takes for an edge to arrive that contracts $G_r$ to $G_{r-1}$ and is
therefore independent of $t_{r'}$ when $r' \ne r$.  But we must ask
whether $t_r$ still has the right exponential distribution---the
complicating factor being that we know the first edge to arrive at
$G_r$ must contract $G_r$ to a specific $G_{r-1}$ and not some other
graph.  To see that this does not matter, let B be the event that first
edge to arrive at $G_r$ is one that creates $G_{r-1}$.  Then
\begin{eqnarray*}
\Pr[t_r \ge t \mid B] &= &\Pr[B \mid t_r \ge t] \Pr[t_r \ge t]/\Pr[B]\\
             &= &\Pr[B]     \Pr[t_r \ge t]/Pr[B]\\
             &= &\Pr[t_r \ge t]
\end{eqnarray*}
since of course, the time of arrival of the edge the contracts $G_r$
has no impact on which of the edges of $G_r$ is the first to arrive.
\end{proofof}

\section{Heuristics and Deterministic Algorithms}
\label{sec:deterministic}

  Until now, we have relied on the fact that the most likely way for a
graph to fail is for some of its near-minimum cuts to fail.  We now
strengthen this argument to observe that most likely, {\em exactly
one} of these near minimum cuts fails.  This leads to two additional
results.  First, we show that the sum of the individual small-cut
failure probabilities is a reasonable approximation to the overall
failure probability.  This justifies a natural heuristic and indicates
that in practice one might not want to bother with the DNF counting
phase of our algorithm.  In a more theoretical vein, we also give a
deterministic PAS for $\fail(p)$ that applies whenever $\fail(p)<
n^{-(2+\delta)}$.  We prove the following theorems.

\begin{theorem}
\labelprop{Theorem}{prop:heuristic}
When $p^c<n^{-4}$ (and in particular when $\fail(p)<n^{-4}$), the sum
of the weak cuts' failure probabilities is a $(1+o(1))$ approximation
to $\fail(p)$.
\end{theorem}

\begin{theorem}
\labelprop{Theorem}{prop:PAS} When $p^c < n^{-(2+\delta)}$ for
any constant $\delta$ (and in particular when
$\fail(p)<n^{-(2+\delta)}$), there is a deterministic PAS for
$\fail(p)$ running in \[(n/\epsilon)^{\exp(O(-\log_n \epsilon))}\] time.
\end{theorem}

We remark that unlike many PASs whose running times are only
polynomial for constant $\epsilon$, our PAS has polynomial running
time so long as $\epsilon = n^{-O(1)}$.  But its behavior when
$\epsilon$ is tiny prevents it from being an FPRAS.

To prove these theorems, we argue as follows.  As shown in
Section~\ref{sec:reliability}, it is sufficient to approximate, for
the given $\epsilon$, the probability that some $\alpha$-minimum cut
fails, where
\begin{eqnarray*}
\alpha &= &1+2/\delta-(\ln \epsilon)/\delta \ln n\\
\end{eqnarray*}
Let us write these $\alpha$-minimum cuts as $C_i$,
$i=1,\ldots,n^{2\alpha}$.  Let $F_i$ denote the event that cut
$C_i$ fails.  We can use inclusion exclusion to write the failure
probability as
\[
\Pr[\cup F_i] = \sum_{i_1}\Pr[F_{i_1}]-\sum_{i_1<i_2}\Pr[F_{i_1}\cap
F_{i_2}]+\sum_{i_1<i_2<i_3}\Pr[F_{i_1}\cap F_{i_2} \cap
F_{i_3}]+\cdots.
\]
Later terms in this summation measure events involving many cut
failures.  We show that when many cuts fail, the graph partitions into
many pieces, meaning a multiway cut fails.  We then argue (using
\ref{prop:r-way vs 2-way simple} or
\ref{prop:r-way failure}) that this is so unlikely that later terms in
the sum can be ignored.  This immediately yields \ref{prop:heuristic}.

To prove \ref{prop:PAS}, we show that for any fixed $\epsilon$ it is
sufficient to consider a constant number of terms (summations) on the
right hand side in order to get a good approximation.  Observe that
the $k^{th}$ term in the summation can be computed deterministically
in $O(m(n^{2\alpha})^k)$ time by evaluating the probability of each of
the $(n^{2k\alpha})$ intersection events in the sum (each can be
evaluated deterministically since it is just the probability that all
edges in the specified cuts fail).  Thus, our running time will be
polynomial so long as the number of terms we need to evaluate is
constant.

\subsection{Inclusion-Exclusion Analysis}

As discussed above, our analyses use a truncation of the
inclusion-exclusion expression for
\[
\Pr[\cup F_i] = \sum_{i_1}\Pr[F_{i_1}]-\sum_{i_1<i_2}\Pr[F_{i_1}\cap
F_{i_2}]+\sum_{i_1<i_2<i_3}\Pr[F_{i_1}\cap F_{i_2} \cap
F_{i_3}]+\cdots.
\]

Suppose we truncate the inclusion-exclusion, leaving out the $k^{th}$
and later terms.  If $k$ is odd the truncated sum yields a lower
bound; if $k$ is even it yields an upper bound.  We show that this
bound is sufficiently tight.  We do so by rewriting the
inclusion-exclusion expression involving particular sets of failed
cuts failing as an expression based on {\em how many} cuts fail.

\comment{
\begin{lemma} 
Let $T_u$ be the event that {\em exactly} $u$ of the events $F_i$
occur.  Then
\[
\sum \Pr[F_{i_1},\ldots,F_{i_k}] = \sum_u \binom{u }{ k} T_u
\]
\end{lemma}
}

\begin{lemma}
\label{lem:incex}
Let $S_u$ be the event that $u$ {\em or more} of the events $F_i$ occur.
If the inclusion-exclusion expansion is truncated at the $k^{th}$
term, the error introduced is 
\[
\sum_u \binom{u-2 }{ k-2} \Pr[S_u].
\]
\end{lemma}
\begin{proof} 
Let $T_u$ be the event that {\em exactly} $u$ of the events $F_i$ occur.
Consider the first summation $\sum F_{i_1}$ in the
inclusion-exclusion expansion.  The event that precisely the events
$F_{j_1},\ldots,F_{j_u}$ occur (that is, the event that cuts
$C_{j_1},\ldots,C_{j_k}$ fail but no others do) contributes to the $u$ terms
$\Pr[F_{j_1}],\ldots,\Pr[F_{j_u}]$ in the sum.  It follows that each
sample point contributing to $T_u$ is counted $u=\binom{u }{ 1}$ times
in the summation.  Thus,
\[
\sum \Pr[F_{i_1}] = \sum_u \binom{u }{ 1} \Pr[T_u].
\]
By the same reasoning, 
\[
\sum \Pr[F_{i_1}\cap F_{i_2}] = \sum_u \binom{u }{ 2} \Pr[T_u],
\]
and so on.  It follows that the error introduced by truncation at term
$k$ is
\begin{eqnarray*}
\lefteqn{
\hspace*{-10em}
\sum_{i_1<i_2<\cdots<i_k}\Pr[F_{i_1} \cap F_{i_2} \cdots \cap
  F_{i_k}]
-
\sum_{i_1<i_2<\cdots<i_{k+1}}\Pr[F_{i_1} \cap F_{i_2} \cap \cdots F_{i_{k+1}}]
+ \cdots}
\\
&= &
\sum_{j \ge k} (-1)^{k-j} \sum_u \binom{u }{ j} \Pr[T_u]\\
&= &\sum_u \sum_{j \ge k} (-1)^{k-j} \binom{u }{ j}  \Pr[T_u]\\
&= &\sum_u \binom{u-1 }{ k-1} \Pr[T_u]
\end{eqnarray*}
Now recall that $S_u$ is the event that $u$ {\em or more} of the $F_i$
occur, meaning that $\Pr[T_u] = \Pr[S_u]-\Pr[S_{u+1}]$.  So we can
rewrite our bound above as
\begin{eqnarray*}
\lefteqn{\sum_u  \binom{u-1 }{ k-1}  (\Pr[S_u]-\Pr[S_{u+1}]) }\qquad\\
&= &\sum_u  \binom{u-1 }{ k-1}  \Pr[S_u]- 
\sum_u  \binom{u-1 }{ k-1}\Pr[S_{u+1}]\\
&= &\sum_u  \binom{u-1 }{ k-1}  \Pr[S_u]- 
\sum_u  \binom{u-2 }{ k-1}\Pr[S_{u}]\\
&= &\sum_u \left( \binom{u-1 }{ k-1}- \binom{u-2 }{ k-1} \right)\Pr[S_{u}]\\
&= &\sum_u \binom{u-2 }{ k-2}\Pr[S_{u}]\\
\end{eqnarray*}
This completes the proof.
\end{proof}

\subsection{A Simple Approximation}

Using the above error bound, we can prove \ref{prop:heuristic}.  Let
$F_i$ denote the event that the $i^{th}$ near-minimum cut fails.  Our
objective is to estimate $\Pr[\cup F_i]$.  Summing the individual
cuts' failure probabilities corresponds to truncating our
inclusion-exclusion sum at the second term, giving (by
Lemma~\ref{lem:incex}) an error of $\sum_{u \ge 2} S_u$.  We now bound
this error by bounding the quantities $S_u$.

\begin{lemma}
\label{prop: many implies big}
If $u$ distinct (2-way) cuts fail then a $\ceil{\log(u+1)+1}$-way cut
fails.
\end{lemma}
\begin{proof}
  Consider a configuration in which $u$ distinct cuts have failed
  simultaneously.  Suppose this induces $k$ connected components.  Let
  us contract each connected component in the configuration to a
  single vertex.  Each failed cut in the original graph corresponds to
  a distinct failed cut in the contracted graph.  Since the contracted
  graph has $k$ vertices, we know that there are at most $2^{k-1}-1$
  ways to partition its vertices into two nonempty groups, and thus at
  most this many cuts.  In other words, $u \le 2^{k-1}-1$.  Now solve
  for $u$ and observe it must be integral.
\end{proof}

\begin{corollary}
\labelprop{Corollary}{prop:prob r fail} If $p^c=n^{-(2+\delta)}$ then
$\Pr[S_u] \le n^{-\ceil{\log(u+1)+1}\delta/2}.$
\end{corollary}
\begin{proof}
Apply \ref{prop:r-way vs 2-way} to the previous lemma.
\end{proof}

Thus, for example, $S_2$ and $S_3$ are upper bounded by the
probability that a $3$-way cut fails, which by \ref{prop:r-way
failure} is at most $n^{-3\delta/2}$.  More generally, all $2^k$
values $S_{2^k},\ldots,S_{2^{k+1}-1}$ are at most
$n^{-(k+2)\delta/2}$.  It follows that the error in our approximation
by the bound of \ref{prop:heuristic} 
is
\begin{eqnarray*}
\sum_{u \ge 2} S_u &\le &\sum_{k \ge 1} 2^k n^{-(k+2)\delta/2}\\ &= &
n^{-\delta} \sum_{k \ge 1} (2n^{-\delta/2})^k\\ &= &
2n^{-3\delta/2}(1+o(1))
\end{eqnarray*}
whenever $\delta>0$.  This quantity is $o(p^c)$, and thus $o(\fail(p))$,
whenever $n^{-3\delta/2} = o(n^{-(2+\delta)})$, i.e. $\delta > 4$.
This proves \ref{prop:heuristic}.

\subsection{A PAS}

We now use the inclusion-exclusion analysis to give a PAS for
$\fail(p)$ when $p^c=n^{-(2+\delta)}$ for some fixed $\delta>0$, thus
proving \ref{prop:PAS}.  We give an $\epsilon$-approximation algorithm
with a running time of $(n/\epsilon)^{\exp(O(-\log_n \epsilon))}$, which is
clearly polynomial in $n$ for each fixed $\epsilon$ (and in fact, for
any $\epsilon = n^{-O(1)}$).

We must eliminate two uses of randomization: in the Contraction
Algorithm for identifying the $\alpha$-minimum cuts, and in the DNF
counting algorithm for estimating their failure probability. 

The first step is to deterministically identify the near-minimum cuts
of $G$.  One approach is to use a derandomization of the Contraction
Algorithm~\cite{Karger:Detcut}.  A more efficient approach is to use a
cut enumeration scheme of Vazirani and
Yannakakis~\cite{Vazirani:EnumerateCuts}.  This scheme enumerates cuts
in increasing order of value, with a ``delay'' of $\Olog(mn)$ per cut.
From the fact that there are only $n^{2\alpha}$ weak cuts, it follows
that all weak cuts (in the sense of Section~\ref{sec:extensions}) can
be found in $\Olog(mn^{1+2\alpha})$ time.

We must now estimate the probability one of the near-minimum cuts
fails.  Let us consider truncating to the first $k$ terms in the
inclusion-exclusion expansion.  From \ref{prop:prob r fail} we know
that $\Pr[S_u] \le n^{-(\log(u+1)+1)\delta/2}$.  It follows from
Lemma~\ref{lem:incex} that for any $k \le \frac13 \delta \log n$, our error
from using the $k$-term truncation of inclusion-exclusion is
\begin{eqnarray*}
\sum_u \binom{u-2 }{ k-2} n^{-(\log(u+1)+1)\delta/2}
        &\le & n^{-\delta/2} \sum_{u\ge k} (u-2)^{k-2} (u+1)^{-\delta (\log
n)/2}\\
        &\le &  \sum_{u \ge k} (u+1)^{k-2-\delta (\log n)/2}\\
        &\le &  \sum_{u \ge k} (u+1)^{\delta(\log n)/3-2-\delta (\log n)/2}\\
        &\le &  \sum_{u \ge k} (u+1)^{-\delta (\log n)/6-1}\\
        &\le &  \int_{u=k-1}^{\infty}  (u+1)^{-\delta (\log n)/6-1} \dee{u}\\
        &= &\frac{k^{-\delta (\log n)/6}}{\delta(\log n)/6}\\
        &= &\frac{n^{-\delta (\log k)/6}}{\delta(\log n)/6}\\
        &= &O(n^{-\delta (\log k)/6})
\end{eqnarray*}
This quantity is $O(\epsilon n^{-(2+\delta)}) = O(\epsilon p^c) =
O(\epsilon\fail(p))$ for some $k=2^{O(-\log_n \epsilon)}$.  It follows
that for an $\epsilon$-approximation we need only evaluate the
inclusion exclusion up to the $k^{th}$ term.  Computing the $k^{th}$
term requires examining every set of $k$ of the $(n/\epsilon)^{O(1)}$
$\alpha$-minimum cuts; this requires
$(n/\epsilon)^{\exp(O(-\log_n\epsilon))}$ time.  This concludes the
proof of \ref{prop:PAS}.

We can slightly improve our bound on $\Pr[S_u]$, which in turn gives better
bounds on $k$.

\begin{lemma}
  If $u$ distinct $\alpha$-minimum cuts fail, then a
  $u^{1/2\alpha}$-way cut fails.
\end{lemma}
\begin{proof}
  Consider a configuration in which $u$ distinct cuts have failed
  simultaneously.  Suppose this induces $k$ connected components.  Let
  us contract each connected component in the configuration to a
  single vertex.  In this contracted graph (before edges fail), the
  minimum cut is at least $c$ (since contraction never reduces the
  minimum cut).  Furthermore, each of the $u$ failed cuts is a cut of
  value at most $\alpha c$, and thus an $\alpha$-minimum cut, in the
  contracted graph.  Since the contracted graph has $k$ vertices, we
  know from \ref{prop:countcuts} that $u < k^{2\alpha}$, meaning that
  $k > u^{1/2\alpha}$.
\end{proof}

However, this serves only to reduce the values of our constants (and
reduce the running time from an exponential to a polynomial dependence
on $1/\delta$).

\section{The Tutte Polynomial}
\label{sec:Tutte}

The {\em Tutte Polynomial} $T(G;x,y)$ is a polynomial in two variables
defined by a graph $G$.  Evaluating it at various points $x,y$ on the
so-called {\em Tutte Plane} yields various interesting quantities
regarding the graph.  In particular, computing the network reliability
$\rel(p)$ is the special case of evaluating the Tutte polynomial at
the point $x=1, y=1/(1-p)$.  Another special case is counting the
number of strongly connected orientations of an undirected graph,
discussed in Section~\ref{sec:orientations}.  Yet another is counting
the number of forests in a graph.  Alon, Frieze, and
Welsh~\cite{Alon:Tutte} showed that for any {\em dense} graph (one
with $\Omega(n^2)$ edges) and fixed $x$ and fixed $y \ge 1$ there is
an FPRAS for the Tutte polynomial.

\subsection{Results}

In this section, we prove the following:

\begin{theorem}
\labelprop{Theorem}{prop:Tutte} For every $y \ge 1$ there is a
$c=O(y\log nxy)$ (in particular, $c=O(\log n)$ for any fixed $x$ and
$y$) such that for all $n$-vertex $m$-edge graphs of edge-connectivity greater
than $c$,
\[
T(G;x,y) = \frac{y^m}{(y-1)^{n-1}}(1+O(1/n)).
\] 
\end{theorem}

Thus, a good approximation can be given in constant time.  Note that
almost all graphs fall under this theorem as the minimum cut of a
random graph is tightly concentrated around $n/2 \gg c$. 

\begin{theorem}
\labelprop{Theorem}{prop:unTutteEasy} 
For every $y>1$ there is a $c = O(y\log nxy)$ such
that there is an FPRAS for $T(G;x,y)$.
\end{theorem}

This theorem is perhaps unsurprising given the previous theorem.  A
slightly more challenging quantity is the ``second order term'' saying
how far a given graph diverges from its approximation in the first
theorem.

\begin{theorem}
\labelprop{Theorem}{prop:unTutteHard}
Let
\[
\Delta T(G;x,y) = \frac{y^m}{(y-1)^{n-1}} - T(G;x,y).
\]
For any fixed $y>1$ and fixed $x$, there is a $c=O(\log n)$ such that
there is an FPRAS for $\Delta T(G;x,y)$.
\end{theorem}

This theorem is stronger than and implies the previous theorem.  When
$\Delta T $ is very close to $0$, $\frac{y^m}{(y-1)^{n-1}}$ accurately
approximates $T$ but approximating $\Delta T$ with small relative
error is harder.

\subsection{Method}
 
Our proofs begin with a lemma of Alon, Frieze, and
Welsh~\cite{Alon:Tutte} (which we have slightly rephrased to include
what is for them the special case of $x=1$):

\begin{lemma}{\cite{Alon:Tutte}}
\labelprop{Lemma}{prop:Alon:Tutte}
When $y > 1$,
\[
T(G;x,y) = \frac{y^m}{(y-1)^{n-1}} E[Q^{\kappa-1}],
\]
where $Q=(x-1)(y-1)$ and $\kappa$ is a random variable equal to the
number of connected components of $G$ when each edge of $G$ fails
independently with probability $p=1-1/y$.  (In the case $Q=0$ (when
$x=1$), we use the fact that $0^r=0$ for $r \ne 0$ while $0^0=1$.)
\end{lemma}

In other words, when $p_r$ is the probability that the graph with random edge
failures partitions into exactly $r$ components, the Tutte polynomial
can be evaluated from
\[
E[Q^{\kappa-1}] = \sum_{k=1}^n p_r Q^{r-1}.
\]
For the remainder of this section, we normalize our analysis by
considering the quantity $T'(G;x,y) = T(G;x,y)\frac{(y-1)^{n-1}}{y^m} =
E[Q^{\kappa-1}]$.  Clearly, any results on relative approximations to
$T'$ translate immediately into results on relative approximations to
$T$.

We begin with an intuitive argument.  From \ref{prop:reliable}, when
$p^c = n^{-(2+\delta)}$ (which happens for some $c=O(\log n)$ for any
fixed $p$) we know $p_r$ is negligible for $r \ge 1$.  Intuitively,
since $p_1 \approx 1$ and all other $p_r \approx 0$, we might as well
approximate $T'$ by $Q$.  Extending this argument, we know that
compared to $p_2$, all terms $p_r$ for $r>2$ are negligible.
Therefore, the error in the approximation of $T'$ by $Q$ is almost
entirely determined by $p_2Q^2$, which we can determine by computing
$p_2$.  

To prove our results formally, we have to deal with the fact that the
term $Q^r$ in the expectation increases exponentially with $r$.  We
prove that the $p_r$ decays fast enough to damp out the increasing
values of $Q^r$.  We also need to be careful that when $Q<0$, the
large leading terms do not cancel each other out.

\subsection{Proofs}

For our formal analysis, instead of the quantities $p_r$, it is more
convenient to work with quantities $s_r$ measuring the probability
that the graph partition into $r$ {\em or more} components.  Note
that $s_1=1$ and $s_2=\fail(p)$.  Since $p_r = s_r-s_{r+1}$, it
follows that
\begin{eqnarray*}
T'(G;x,y) &= &\sum_{r=1}^n p_r Q^{r-1}\\
&= &\sum_{r=1}^n (s_r-s_{r+1}) Q^{r-1}\\
&= &\sum_{r=1}^n s_rQ^{r-1}-\sum_{r=2}^n s_r Q^{r-2}\\
&= &1 + \sum_{r = 2}^n s_r (Q^{r-1}-Q^{r-2})\\
&= &1 + (Q-1) \sum_{r=2}^n s_r Q^{r-2}  
\end{eqnarray*}
\ref{prop:Tutte} will follow directly from the last equation if we can
show that the trailing term $(Q-1) \sum_{r=2}^n s_r Q^{r-2} = O(1/n)$.
\ref{prop:unTutteHard} will follow if we can give an FPRAS for
$\sum_{r=2}^n s_r Q^{r-2}$.  The fact that the value of this sum is
$o(1)$ (\ref{prop:Tutte}) means that the FPRAS for it immediately
yields an FPRAS for $T'$, thus proving \ref{prop:unTutteEasy}.

To prove these results, first consider the case $x = 1$.  In this
case \mbox{$Q=0$}, meaning $Q^{r-2}=1$ for $r=2$ and $0$ for $r>2$.  Thus
$T'(G;x,y) = 1-s_2 = 1-\fail(p) = \rel(p)$.  We have already seen in
\ref{prop:reliable} that whenever $p^c = n^{-(2+\delta)}$, the probability
that the graph becomes disconnected is at most
$n^{-\delta}(1+2/\delta)$.  This is certainly $O(1/n)$ if $\delta \ge
1$, meaning $\rel(p) = 1-O(1/n)$.  But this in turn is true when $p^c <
n^{-3}$, i.e.
\[
c > 3\ln n/\ln(y/(y-1)) =O(y\ln n).
\]
This proves \ref{prop:Tutte} for $Q=0$.  On the other hand,
\ref{prop:unTutteHard} simply claims that there is an FPRAS for
$1-\rel(p) = \fail(p)$, which is what Section~\ref{sec:reliability}
showed.  Finally, \ref{prop:unTutteEasy} says that when $\fail(p)$ is
small, we can approximate $\rel(p)$ (by approximating $\fail(p)$).

We now generalize this argument to the case $x > 1$.  
To derive the appropriate lower bound on $c$, we state two criteria
that will we need in our analysis.  First, we require $c$ to be such
that $p^c = n^{-(2+\delta)}$ for some $\delta > 1$.  Equivalently, we
have $1\le\delta = -\log(n^2p^c)/\log n$. Second, we require that
$Q<\frac14n^{\delta/4}$.  Plugging in for $\delta$, we find the
equivalent requirement
\begin{eqnarray*}
Q & < &\frac14n^{\delta/4}\\
&= &\frac14 (n^2p^c)^{-1/4}\\
(4Q)^4 &< &1/n^2p^c\\
n^2(4Q)^4 &< &(\frac{y}{y-1})^c\\
\ln(256Q^4n^2)/\ln(1+\frac{1}{y-1}) &< &c
\end{eqnarray*}
This is true for some $c=O(y(\ln nQ))=O(y\ln nxy)$ as
claimed.

Given the above relations between $Q,n,$ and $\delta$, we can use
\ref{prop:r-way vs 2-way}.  Since $p^c = n^{-(2+\delta)}$, we deduce
that $s_r \le n^{-r\delta/2}$.  Since $Q < \frac14n^{\delta/4}<
\frac12n^{\delta/2}$ we find that
\begin{eqnarray}
\sum_{r=r_0}^n s_r Q^{r-2} &\le &Q^{-2}\sum_{r \ge r_0}
(Qn^{-\delta/2})^r \label{eq:mysum}\\
\nonumber &\le & Q^{-2}(Qn^{-\delta/2})^{r_0}/(1-(Qn^{-\delta/2})^{r_0})\\
\nonumber &\le & Q^{-2}(Qn^{-\delta/2})^{r_0}/(1-\frac1{2^{r_0}})\\
\nonumber &\le & 2 Q^{-2}(Qn^{-\delta/2})^{r_0}\\
\end{eqnarray}
Our results follow from this bound.  First, taking $r_0=2$, we find that
the error in approximating $T'(G;x,y)$ by $1$ is at most
\[
2n^{-\delta} = o(1).
\]
This proves \ref{prop:Tutte}.  

To prove \ref{prop:unTutteHard}, note that the leading term in the
 summation~(\ref{eq:mysum}) is $s_2 \ge n^{-(2+\delta)}$.  We can therefore
estimate the sum to within relative error $O(\epsilon)$ by evaluating
summation terms up to summation index $r_0$ where
$(Qn^{-\delta/2})^{r_0} \le \epsilon n^{-(2+\delta)}$. 
Since the left hand side decreases exponentially in $n$ as a function
of $r_0$, we can achieve this error bound by taking
\[
r_0 = O(\log_n (n^{2+\delta}/\epsilon)) = O(1+\log_n 1/\epsilon).
\]
In other words, we need only need to determine $O(1-\log_n\epsilon)$
terms in the summation.  This in turn reduces to determining the
quantities $s_r$ appearing in those terms.

We cannot find the $s_r$ exactly. However, for an
$\epsilon$-approximation, it suffices to approximate each relevant
$s_r$ to within $\epsilon$.  We can do so using the algorithm of
\ref{prop:multiway}.  The running time of this algorithm for
estimating the $r$-way failure probability to within $\epsilon$ is
$(n^r/\epsilon)^{O(1)}$.  We have argued above that we only need to
run the algorithm for $r \le r_0 = O(1-\log_n \epsilon)$.  It follows
that the running time of our algorithm is
$n^{O(1-\log_n\epsilon)}/\epsilon^{O(1)}= (n/\epsilon)^{O(1)}$, as
required.  This proves \ref{prop:unTutteHard}.

Finally, we consider the case $x<1$.  Our argument is essentially
unchanged from before.  We need to be slightly more careful because
our sum is now an alternating sum, which means that the leading terms
are a good approximation only if they do not cancel each other out.
To see that such cancelling does not occur, note that the first term
has value $s_2=n^{-(2+\delta)}$, while the remaining terms (by the
analysis above) have total (absolute) value $O(n(Qn^{-3\delta/2}))$.
If we choose $n$ large enough that $Q<\frac14 n^{\delta/4}$, then this
bound is $O(\frac14 n^{-5\delta/4}) < \frac14 s_2$ for $\delta > 4$,
so the remaining terms do not cancel $s_2$.

\section{Conclusion}

We have given an FPRAS for the all-terminal network reliability
problem and several variants.  In the case of large failure
probability, the FPRAS uses straightforward Monte Carlo simulation.
For smaller failure probabilities, the FPRAS uses an efficient
reduction to DNF counting or a less efficient deterministic
computation.  An obvious open question is whether there is also a
deterministic PAS for the case of large failure probabilities.
Another is whether there is also an FPRAS for $\rel(p) =1-\fail(p)$,
the question being open only for the case $\rel(p)$ near $0$.

This work has studied probabilistic {\em edge} failures; a question of
equal importance is that of network reliability under {\em vertex}
failures.  We are aware of no results on the structure of minimum
vertex cuts that could lead to the same results as we have derived
here for edge cuts.  In particular, graphs can have exponentially many
minimum vertex cuts.  The same obstacle arises in {\em directed}
graphs (where we wish to measure the probability of failing to be
strongly connected).

Although the polynomial time bounds proven here are not extremely
small, we expect much better performance in practice since most graphs
will not have the large number of small cuts assumed for the
analysis.  Preliminary experiments~\cite{Karger:ImpRel} have suggested
that this is indeed the case.

\bibliographystyle{siam}
\bibliography{me}

\end{document}